# The energy-population conundrum and its possible solution

Francesco Meneguzzo,[1*] Rosaria Ciriminna,[2] Lorenzo Albanese,[1] Mario Pagliaro[2*]

[1]*Istituto di Biometeorologia, CNR, via G. Caproni 8, 50145 Firenze, Italy;* [2]*Istituto per lo Studio dei Materiali Nanostrutturati, CNR, via U. La Malfa 153, 90146 Palermo, Italy*

**Abstract**
Oil prices above $100/barrel values have proven unaffordable for the world economy, while lower prices have proven unaffordable for unconventional oil sources, resulting in a frantic price swing since 2007-2008. We identify and combine for the first time the competing dynamics of oil price, economic growth and extraction costs in a single model aiming to evaluate the near-term consequences of these dynamics onto forthcoming oil supply. Policies able to cope with the consequences of the resulting energy scenario are suggested in the conclusions.

**Keywords**: oil shortage; solar energy; renewable energy; tight oil; electrification

*Corresponding Authors*

Dr. F. Meneguzzo
Istituto di Biometeorologia, CNR
via G. Caproni 8
50145 Firenze (Italy)
E-mail: francesco.meneguzzo@cnr.it

Dr. M. Pagliaro
Istituto per lo Studio dei Materiali Nanostrutturati, CNR
via U. La Malfa 153
90146 Palermo (Italy)
E-mail: mario.pagliaro@cnr.it

# 1. Introduction

In the course of the last fifteen years (2000-2015), one of the most impressive changes in the energy scenario has been the accelerated deployment of renewable energy sources at the global scale, so much that the perspective of worldwide wind and solar electricity supply is no longer an utopic exercise [1 2].

Along with wind energy, easily scalable and versatile solar photovoltaics has been massively adopted, first in Europe and now across the world, to become the energy technology with the fastest rate of adoption, and chances for further improvement [3]. Almost concomitantly, since 2009 another large change occurred in the exploitation of new energy sources, namely, the successful extraction and marketing of unconventional hydrocarbons, mainly shale gas and shale oil (the latter hereinafter also referred to as "tight oil"), relying on a combination of improved and massively deployed extraction techniques, such as horizontal drilling and fracking. Started to be seriously explored in the US in 2007-2008, the exploitation of these resources evolved into massive production in mid-2011, allowing that country to attain self-sufficiency in gas production in dramatically short time [4].

In detail, more than 4 million barrels per day (b/d) of tight oil added to the global production of petroleum (roughly 5% of the world production), with the most noticeable increase occurring after 2010, driven by sustained high oil prices [5]. Eventually, in 2015 the global oil supply had increased by more than additional 9 million b/d, challenging the very concept of "peak oil", namely the theory predicting that the world production of crude oil should have already reached a maximum (peak), to gradually decline until vanishingly levels [6].

In the spring of 2015, however, a possible imminent decline of tight oil successfully extracted during the previous four years was reported [7]. In detail, the crude oil output from the seven most productive shale formations in the US started declining for the first time since the beginning of the large-scale intensive exploitation of tight oil in 2011 (the predicted output fall was around 2,100 barrels b/d for April 2015; 56,000 b/d for May, and so on). Early signs of production decline arose in March 2015 weekly data [8], following a steep fall of the oil rigs count since November 2014 [9], as the oil price was falling from about $110/b on June 2014, to about $48/b in January 2015. After a deep minimum of $31/b in January 2016, prices recovered to about $50/b as of June 2016, since then revealing relatively stable.

Certain energy analysts have argued that once the oil price will rebound towards $100/b the uncompleted wells will be resumed, and oil rigs count shall increase again. Others, conversely, rule out any chance for production recovery in light of the demonstrated sensitivity of the tight oil industry to the oil price [10].

Due to its intrinsically high extraction costs, the "well-head" energy return on energy invested (EROI) for tight oil is in the 1.5-2 range [11], namely about one tenth of the 11-20 set of values for conventional crude oil [12,13]. Unconventional crude oil extracted from oil sands (those fields having contributed substantially to the production increase in recent years, even though lesser than tight oil), or from ultra-deep waters, show EROI ranges comparable with tight oil from shale deposits [13]. Specifically, the reasons for such a low EROI are the use of large quantities of energy from the same shale formation during the process (including liquefaction of the originally solid kerogen), as well as the need for a huge number of drills, thus of oil rigs, in order to keep pace with the rapidly declining extraction rates of single wells [14].

More recently, Reynolds applied advanced modeling of the extraction rate and depletion of conventional and unconventional oil resources [15]. He concluded that a significant shortage of oil supply will shortly occur, as a consequence of the synergistic combination of the ever growing oil demand of the global economy, and its inability to pay the high oil prices needed for the economic viability of unconventional resources.

Using modern non-linear scaling theories, now we combine the competing dynamics of population and economic growth with oil supply and price, aiming at evaluating the resulting near-term consequences onto global economic growth. We conclude suggesting energy policies able to cope with the undesirable consequences of the identified trend.

## 2. Datasets

The datasets and the respective sources listed in Table 1 include demographic, energetic and financial information at annual and monthly frequency, on global as well as on single country scale.

Table 1. Datasets and sources

| Dataset | Period and frequency | Unit | Source* |
|---|---|---|---|
| World population | 1950-2015 Annual | Billions | UN [16] |
| World gross domestic product (current US$) | 1960-2015 Annual | Trillion US$ | World Bank [17] |
| Broad money Industry value added | 1960-2015 Annual | % of GDP | World Bank [17] |
| World energy consumption by source | 1965-2015 Annual | MTOE | BP [18] |
| Brent oil price | 1965-2015 Annual | $ per barrel | BP [18] |
| | 1987-2016 (May) Monthly | | EIA [8] |
| World and regional crude oil production | 1987-2016 (May) Monthly | Million barrels per day | JODI [19] |

*GDP = Gross Domestic Product; MTOE = Million Tons of Oil Equivalent; EIA = Energy Information Administration; UN = United Nations; JODI = Joint Organizations Data Initiative; BP = British Petroleum

## 3. Population, wealth and energy consumption

Following non-linear scaling theories recently introduced in socio-economic studies [20], the world gross domestic product (GDP) was modeled as a power function of the global population. Figure 1 shows the power law fitting the world GDP data series, plotted against the population, formally represented by Equation 1, which explains more than 99% of the sample variance:

$$W\_GDP = (0.0077 \pm 0.0008) W\_Pop^{(4.69 \pm 0.06)} \qquad (1)$$

where W_GDP is the world GDP, in units of trillion current US dollars, and W_Pop is the global population, in units of billions.

Simple differentiation of Equation 1 provides the expected year on year GDP growth rate d(W_GDP) as described by Eq.2, where d(P) is the population growth rate, in the same units of Eq.1:

$$d(W\_GDP) = 0.0361 \, W\_Pop^{3.69} d(P) \qquad (2)$$

Relationships similar to Eq.1 can be proved to hold for single countries and economic areas such as the European Union, Germany, France, Italy, or the US, showing that the power law fitting of the GDP as a function of population has been an ubiquitous feature of the world economic development at least since 1960, with a specific exponent of the power law assigned to each economic area.

Nevertheless, a deviation from the exponential relationship is noted during the last two years of the series, *i.e.* 2014 and 2015, with a sudden downturn of GDP in 2015.

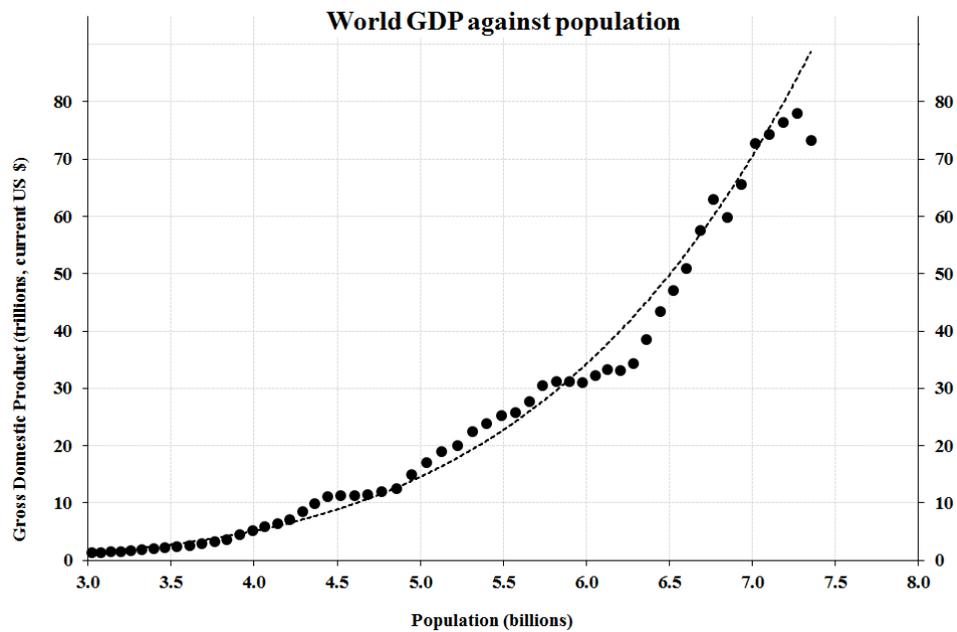

**Figure 1**. Power law fitting the gross domestic product (GDP) as a function of the global population, during 1960-2015.

As shown in Figure 2(a), the total energy consumption (TEC) has steadily increased, literally fueling the growth of global wealth, although at a pace only proportional to the overall population (98% of variance explained by a linear fit).

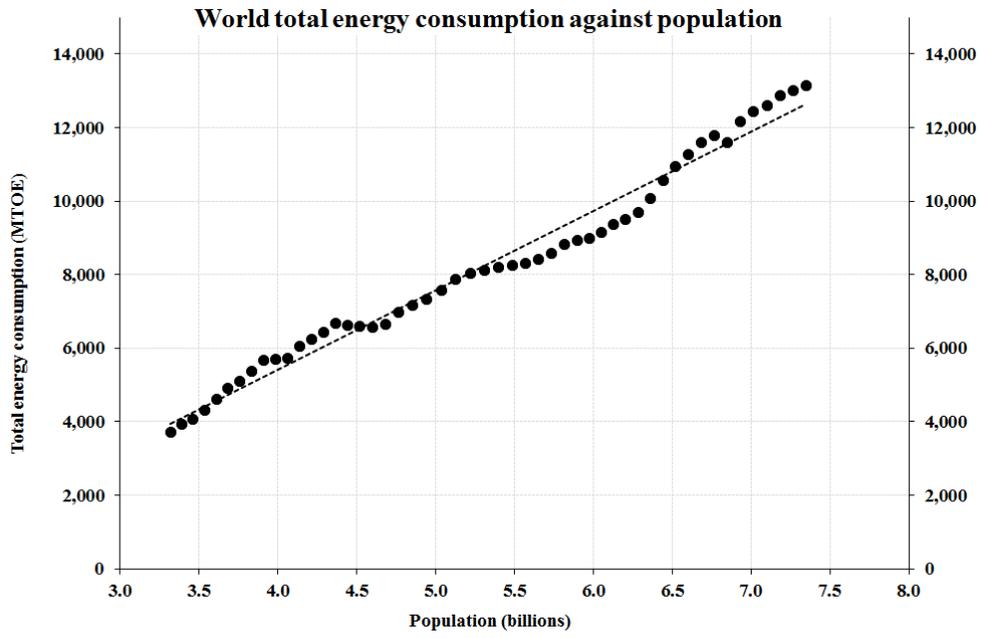

(a)

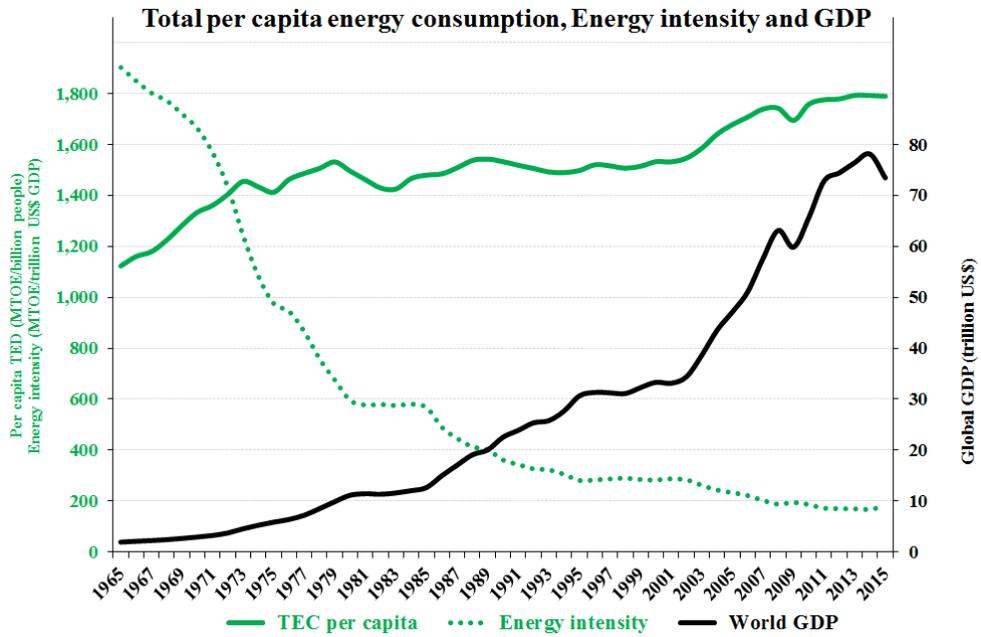

(b)

**Figure 2**. World total energy consumption (TEC) as a function of population (a), and time series of per capita TEC, energy intensity and GDP (b), during the 1965-2015 period.

In detail, Figure 2(b) shows that the specific (per capita) TEC has enjoyed two distinct periods of growth since 1965; the first one in 1965-1973, and the second during 2001-2008, with GDP growth rate accelerating in the latter period and pointing to an intrinsic crucial role of the population size. However, in the last three years of the series (2013-2015), the individual (per capita) TEC was levelling, with the GDP growth first slowing down and eventually reversing with a worrisome drop in 2015.

Again, in Figure 2(b), the energy intensity (energy consumed per unit GDP) shows an impressive decline that closely follows an exponential decay trend, although an apparent levelling is visible in the latter years of the series. As shown in Figure 3(a), the energy mix has profoundly changed during the 50 years period under consideration, most notably with the gradual decline of the oil contribution, particularly since late 1970s, partly replaced by natural gas and coal. Nuclear energy, remarkably, has covered a tiny fraction of the global consumption and, even more important, it peaked in 2006 at 635 MTOE (5.6% of the total). In 2015, the above figures for the nuclear source reduced to 583.1 MTOE and 4.4%, respectively.

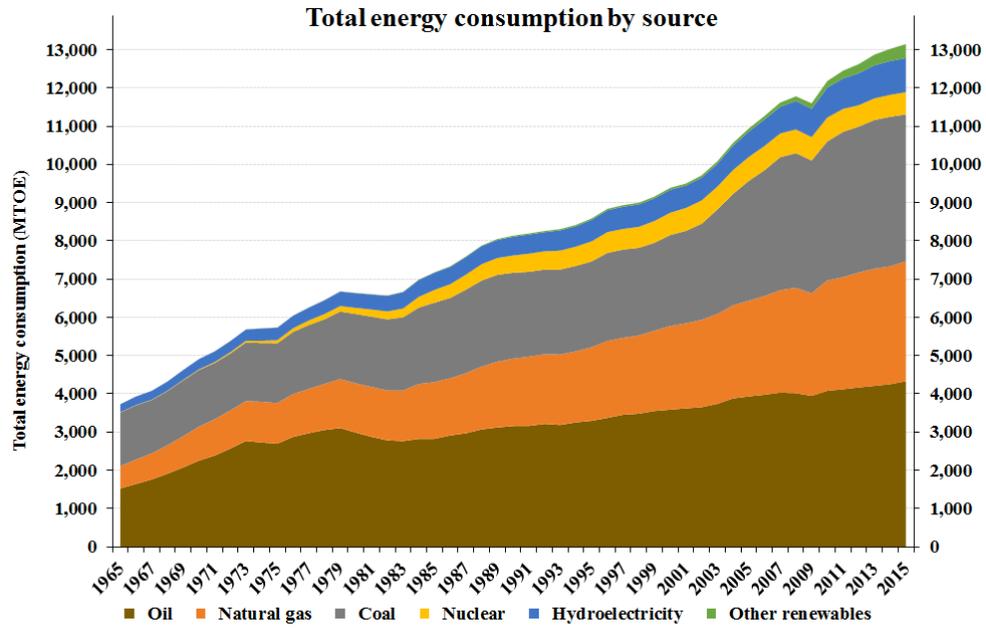

(a)

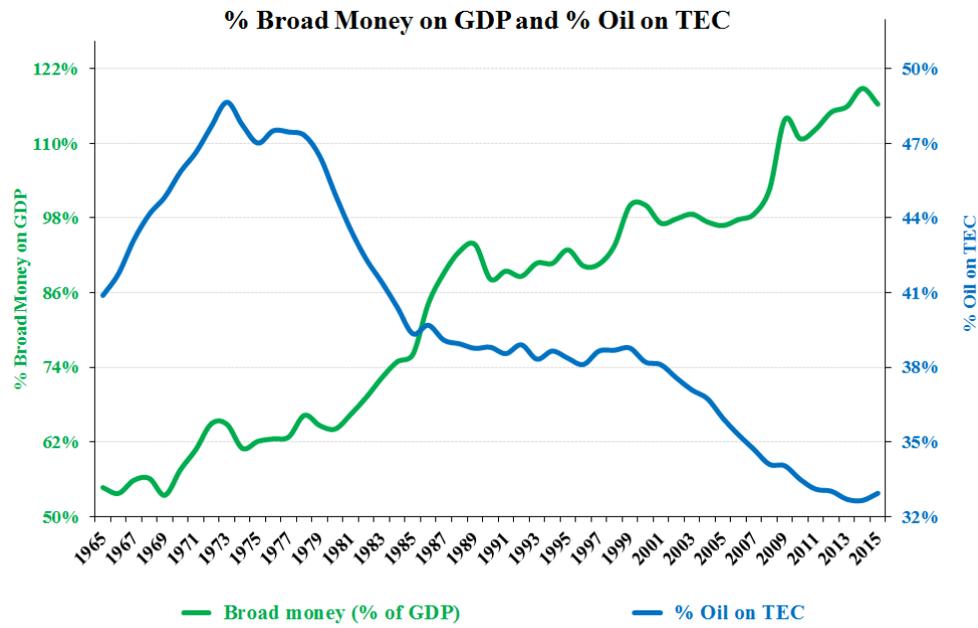

(b)

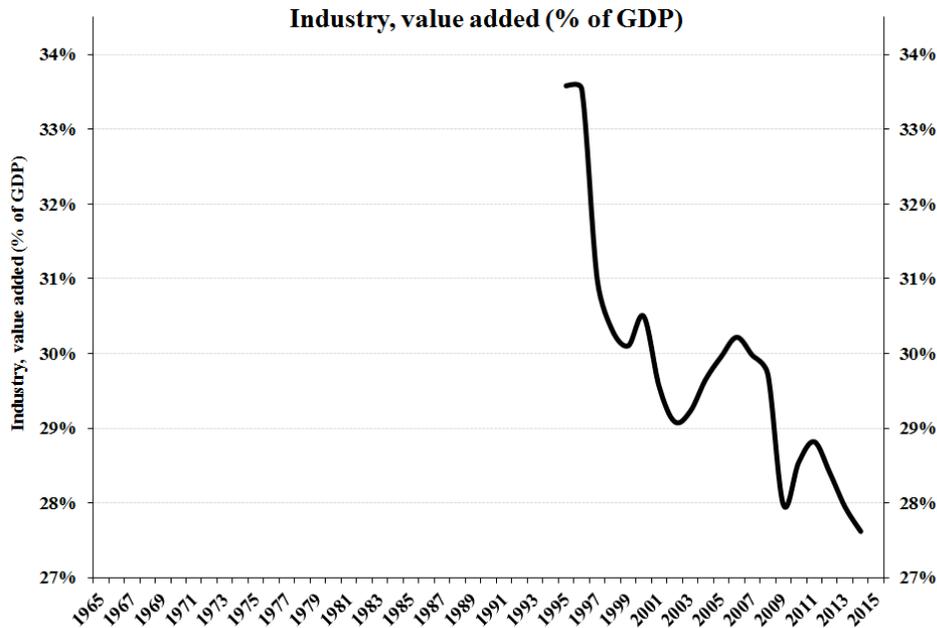

(c)

**Figure 3**. Energy consumption by source (a), time series of broad money and percentage oil on TEC (b), industry value added as percentage of GDP (c), during 1965-2015.

Figure 3(b) shows that, at the same time, the global amount of money (% of broad money on GDP), faithfully representing the amount of debt [21], started a relentless growth, accelerating just during the steeper phases of oil decline in the energy mix, *i.e.* during most of the 1980s and in the early 2000s, hinting to the attempt by Governments and financial institutions to buy the current growth from the future amid a tightening resource basis useful for growth [22–24].

Supporting the latter point, Tverberg recently drew attention to the role of the energy concentration and transportability as key parameters to define the hierarchy of energy sources, as well as a major driver of economic growth,[25]. Oil is by far the most concentrated energy source applicable to virtually all end uses.

Hence, it may not be surprising to learn that its downward trend in the energy mix has led to substantial economic difficulties. As a clue to such difficulties, Figure 3(c) shows that industry value added as a percentage of GDP (global data available only during 1995-2014) has been falling quite rapidly.

This evidence suggests that most of GDP growth, at least in the last two decades, can be ascribed more to the increase in debt than to real wealth generation; and that the concurrent effect of the declining EROI of traditional fuels has played a substantial role in the dynamics identified [26]. As a consequence, the chance of future economic growth matching the current trajectory of the human population is inextricably bound to the wide and growing availability of highly concentrated energy sources enjoying broad applicability to energy end uses.

## 4. Oil price and wealth growth

Sufficient clues exist that the oil price rising above a certain threshold has been a concurrent cause of recurrent slowdowns and recessions of the global economy [13, 22, 27, 28]. In particular, spikes in oil price were deemed responsible for ten out of eleven recessions in the US since World War II [29]. Higher cost of oil would act as an highly effective "tax" inhibiting economic growth [30].

In the frame of the population-driven, energy-assisted, simple model of wealth growth shown by Eq.1, Figure 1 and Figure 2(a), an unbiased index of anomalies affecting the global economy is the difference between year on year growth rates of the GDP computed from real data and from the simple model described by Equation 2.

Figure 4 shows the series of such GDP growth anomalies in the period 1961-2015 along with inflation-adjusted oil price since 1955, with the four main events (labelled A, B, C and D) allegedly connecting in a causative manner the spikes in oil price with subsequent global economic slowdown.

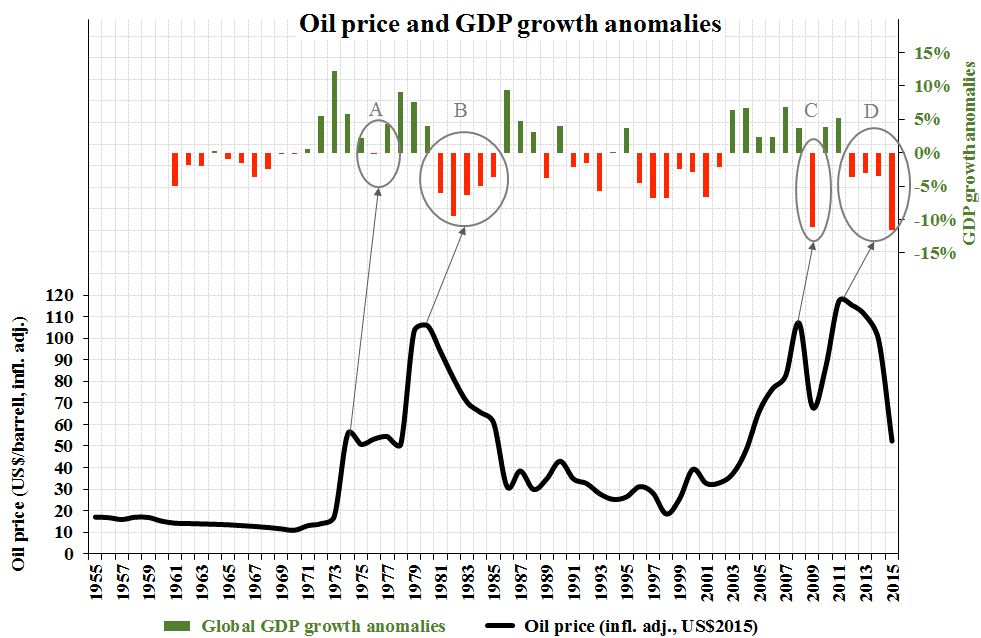

**Figure 4**. Time series of inflation-adjusted oil price and global GDP growth anomalies. Event labelled A to D mark economic slowdown or recession linked to high oil prices.

The "A" event occurred after the first "oil shock" in 1973-1974 (maximum yearly average price about $56/b). Its effect was short-lived, being limited to 1976 (but occurring during a relatively strong positive deviation). The "B" event occurred after the second oil shock in 1979-1980 (maximum yearly average price about $106/b), with a significant and sustained slowdown during the years 1981-1985. The "C" event occurred immediately after the strong price spike of 2008 (average price $107/b), following a period of sustained growth taking place since 2003, resulting in the earnest recession of 2009. Finally, following high prices culminated in 2011 with the $117/b price, the "D" event features a series of negative anomalies of GDP growth in 2012-2015, with 2015 marked by the strongest negative anomaly during the whole 1961-2015 period.

Comparing Figure 4 with Figure 2(b), it is immediate to notice that all such events (A-D) were in synchrony with both dips or stagnation of per capita total energy consumption, and steep falls of oil price.

## 5. A natural oil price?

Figure 5 shows the monthly Brent oil prices from January 2002 through June 2016, encompassing events "C" (monthly peak at $146/b of July 2008) and "D" (monthly peaks around $127/b on April 2011 and March 2012).

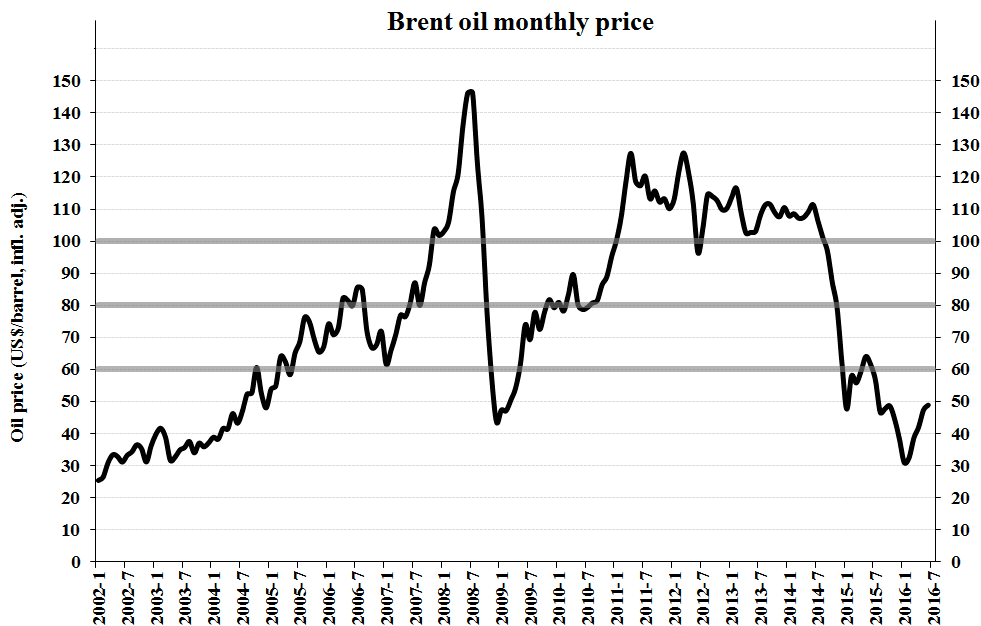

**Figure 5.** Series of monthly average inflation-adjusted Brent oil price, (Jan 2002 - Jun 2016).

In each of those events, the residence time of oil price above the $100/b and $80/b thresholds are very similar, while during the steep fall following a price spike, the threshold of $80/b is crossed very quickly, until a bottom price somewhat lower than $60/b, around the price levels of the 1990s, is reached.

Here we advance the hypothesis that an underlying force originated in the global interconnected community (or, if preferred, Smith's invisible hand) works to steer down the oil prices to levels well lower than $60/b (in units of 2013 US dollar currency), in order to ensure the stability of the GDP "natural" growth rate, as derived by Eq.2. Supporting this hypothesis, in 2014 Murphy was noting that the average oil price during periods of economic growth over the past 40 years was under $40 per barrel, while the average price during economic recessions was under $60 per barrel [13].

The above observations corroborate the idea that oil has played a key role to sustain the GDP natural growth rate. Furthermore, the sustainability of additional debt, as represented by % of broad money on GDP in Figure 3(b), requires that the oil fraction in the energy mix should climb again to approximately 40% from current 33% value, while the oil price should not exceed a threshold located somewhat between $40/b and $50/b, or possibly even lower. In alternative, another universally applicable energy source, similarly abundant, concentrated and cheap, should replace petroleum.

6. Global oil supply

Figure 6(a) shows the worldwide monthly crude oil production series in million b/d, thus excluding other products, such as liquids produced from coal and gas, Orimulsion, biofuels such as biodiesel and ethanol, as well as other hydrocarbons, contributing to the overall oil supply. Moreover, the global crude oil supply is partitioned into outputs from the four biggest producers (USA, Russia, Saudi Arabia and China), and the rest of the world. Crude oil production in US has been especially relevant because almost 70% of the increase in worldwide crude oil supply in the period 2005-2014 (about 5 million b/d in 9 years) has been due to output growth from US. Indeed, while in June-August 2011 the US crude oil output was around 5.5 million b/d, in April 2015 it achieved an astonishing volume of 9.7 million b/d, namely very few hundred thousand barrels lower than the historical peak occurred in November 1970.

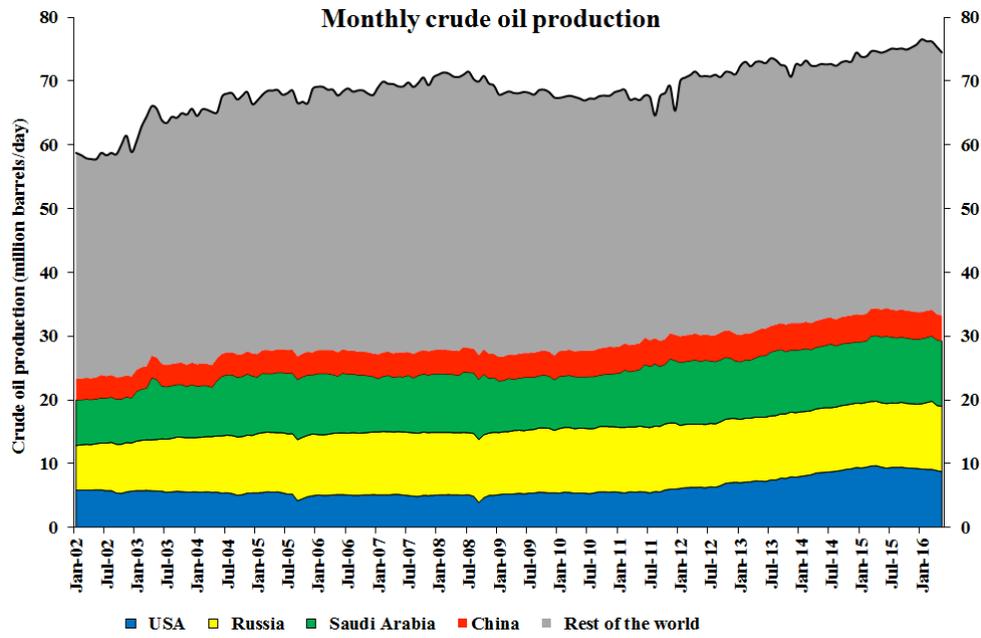

(a)

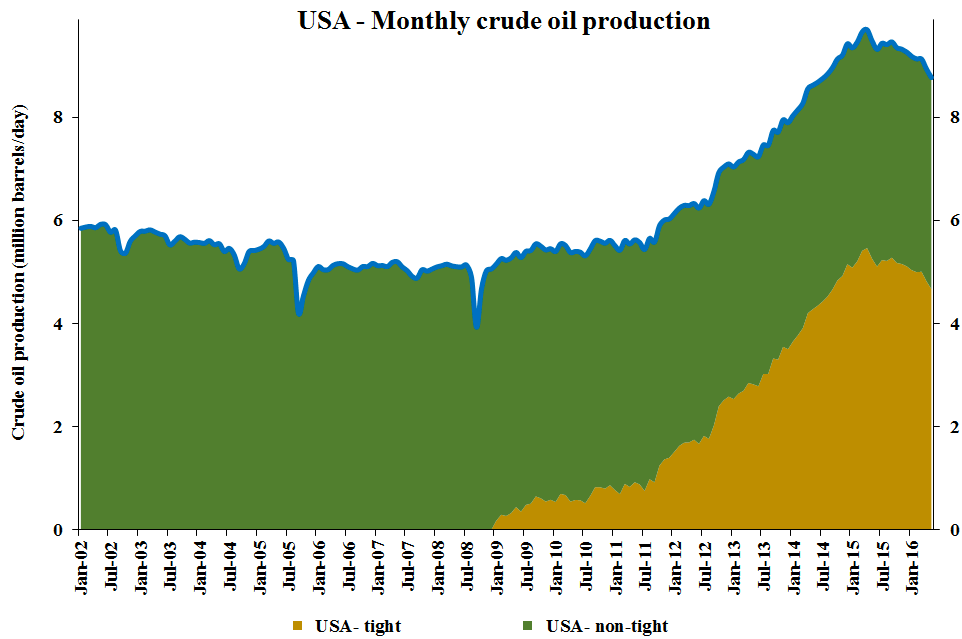

(b)

**Figure 6.** Series of monthly averaged oil production: global (a) and USA (b), Jan-2002 to May-2016 (units in million barrels daily).

Assuming January 2009 as the start date of significant tight-oil extraction, and considering the monthly US crude oil output since September 1, 1992, the long-term trend of the US non-tight crude oil output (production $P_{nt}$ in units of million b/d), before tight oil exploitation began, can be represented as a linear decreasing function of time, with around 90% of variance explained by Eq.3:

$$P_{nt} = (-0.119 \pm 0.002)t + (6.92 \pm 0.01) \qquad (3)$$

where t is time in units of years starting with t=0 on September 1, 1992. Assuming such trend has continued after 2008, a rough estimate for the tight oil output can be obtained after subtracting the linear trend represented by Eq.3 from the complete series.

As mentioned above (Section 1), the tight economics of shale oil extraction indicates that roughly one-third of current production could be uneconomical at oil prices around $60 per barrel. Moreover, most tight and other unconventional oil resources extracted in the US, Canada, Brazil and Mexico would require prices close to $80/b [31], or even higher [13, 29].

Indeed, Figure 6(b) shows evidence that the exploitation of tight oil became significant in the overall US output during 2011, at the same time of the fast increase of oil prices persisting in the course of the sustained high prices of 2011-2014. Driven by the peak of tight oil, a sharp peak in US total crude oil production occurred in March-April 2015. The unbalance between overall costs of US tight oil production and price of oil, which began to declines a few months earlier to cross the $60/b on December 2014 (Figure 5), is self-evident.

It is also remarkable that the pattern of US tight oil production fully confirmed the arguments advanced by Murray and Hansen [29] and, independently, by Hughes [14], based on the very different nature of conventional and shale oilfields, with regards to exploration and extraction techniques, patterns of growth and depletion.

In a recent study, Wang and coworkers performed a rigorous modeling approach to forecast oil production in China, separating the conventional and unconventional (*i.e.* tight oil) fractions for the very first time. Their best estimate locates the peak in the year 2020, with narrow associated uncertainty in timing [32].

In brief, unless conventional production from major producers is substantially increased, it is unlikely that unconventional resources such as tight oil, Canadian and Venezuelan oil sands, or biofuels, can help keeping the pace with the global oil need associated to the "natural" growth rate. Indeed, Canadian oil sands are assessed uneconomical at oil prices even higher than those typical of US tight oil [13, 31], while the heavily subsidized biofuels are losing momentum and substantially plateauing in output [27].

In a trend exactly opposite to what could be expected, lately major crude oil producers have been extracting oil around their peak capacity despite the unfavorable market prices, thus shortening the time to peak and subsequent depletion [33]. A very recent model based on past production data for conventional and unconventional oil suggests that only certain Middle East countries might maintain their current production levels during the next decade [34]; while all other producing regions will experience decline already between 2015 and 2020. As pessimistic as it may seem, the latter study omits to consider that even Middle East crude oil is not becoming more available outside the production area, mostly due to the fast growth of population and domestic consumption in those countries during the last 30 years.

## 7. Alternative sources and energy end-use

If global population will keep growing along the current trajectory, in 2016-2025 about 800 million people will add to current population. Correspondingly, according to the linear relationship shown in Figure 2(a), in 2025 the total energy consumption should increase by about 1,700 million tons of oil equivalent (MTOE) per year over the current level, in order to feed the global "natural" growth of the GDP (wealth). Even to keep the oil fraction in the energy mix at the current level around 33%, more than 11 million b/d should be added to current production levels.

On the other hand, in order to achieve the desirable threshold of 40% for the oil fraction in the energy mix, additional 32 million b/d would be needed. Such figures should be compared with the <10 million b/d added during the 2005-2015 decade when the oil price was high (averaging around $90/b), including all liquids, out of which crude oil only accounted for 7.4 million b/d. Recalling that most of the contribution to the increase in crude oil supply came from US tight oil, which is currently declining being economically unsustainable at price levels compatible with global economy growth, perspectives could be even more uncertain.

Figure 3(a) shows that natural gas and, even more, coal, have emerged as candidates to replace oil as major sources in the energy mix. Natural gas, (leaving aside liquefaction meant to increase the concentration and ease transportability in the form of liquefied natural gas, consuming at least 25% of the internal energy [35]), has limited concentration and requires the deployment of long pipelines such as those connecting Russia to Europe or Libya and Algeria to Italy. Though growing, the global production of natural gas is insufficient to compensate for the production decline of oil and coal [5].

As to coal, its practical applicability only to power generation represents the basic problem. The increase of its share in the energy mix has been chiefly due to impressive growth of lignite utilization in China (and to a lesser extent in India too). Yet, in 2009 China became a net coal importer, with recent assessments placing the peak of lignite domestic production between 2025 and 2030 [36].

Uranium, as the source of nuclear power, enjoys by far the greatest energy concentration and its transportation is not an issue. Beyond very limited direct applications to the transportation sector such as in nuclear reactors powering icebreakers, military submarines and ships, its only practical use for society lies in power generation, requiring a shift to electrification of energy end uses to achieve universal applicability. Although such shift is indeed possible and even desirable, relevant problems for uranium arise from its availability and cost [37]. Production is forecast to peak in the second half of 2010s, followed by a slow decline up to 2025 and steeper afterwards, revealing insufficient even to feed existing and already planned nuclear power plants [38].

A recent comprehensive study of Cartelle Barros and coworkers [37] shows that the economics of the most representative renewable energy sources (RES) such as high-temperature solar thermal (ST), onshore wind, solar photovoltaic (PV) and small hydropower, are entirely comparable with that of oil, natural gas and coal. Limited to solar PV, the above results are likely to derive from the already significant EROI of all its technological variants even in mid-latitude areas [2], and to the steep energy learning curve of PV technology, making its deployment increasingly competitive with conventional sources at latitudes as high as 65° [39]. Overall availability is not an issue for the solar and wind sources [40]; but while deployment of high-temperature ST is profitable only at relatively low latitude, high insolation areas, solar PV can be deployed over a much wider portion of the global world [40, 41].

As to the wind source, the geographical distribution of its availability is more dispersed than for solar PV, with its seasonal and solar day availability almost complementary to that of insolation. Actually, in the case of RES, the issues of availability, concentration and applicability are tightly interlinked. Since their applicability is practically limited to power generation, the electricity transmissibility over long distances makes original concentration a minor issue, provided that availability is ensured around the clock and energy end uses can be fulfilled by electricity.

A 100% grid penetration of intermittent wind, water and solar generated power for all purposes with electrification of virtually all energy end uses has been lately advocated by Jacobson and co-workers for the US [42], and by us in the case of Italy [43]. In such an electricity-powered world based upon intermittent renewable sources, a significant fraction of energy will be distributed across all-battery vehicles [44]. Along with electrification of the energy end uses, breakthrough advances in storage technologies such as that being developed by Zhang and co-workers based on enzymatic $CO_2$ and renewable electricity fixation in carbohydrate polymers [45] will be crucial to complete the energy transition, and replace oil and other fossil fuels in the energy mix by RES [2, 40].

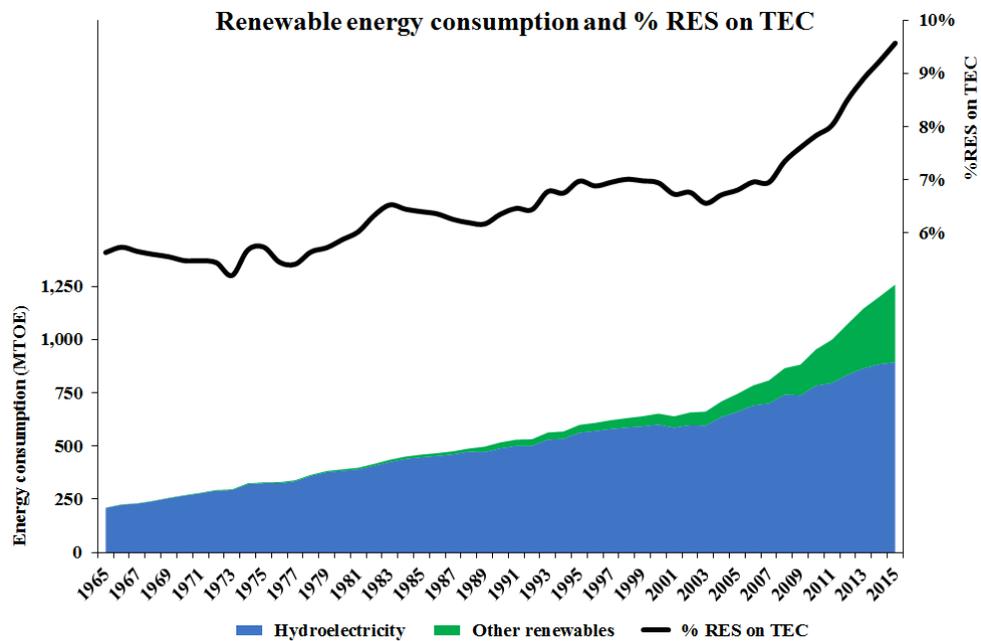

**Figure 7.** Renewable energy consumption, partitioned between hydroelectricity and other sources, and its percentage on total consumption, during 1965-2015.

In brief, the share of RES over the total energy consumption will grow at significantly faster pace than so far. Figure 7 shows that wind and solar PV [3] (most of "other renewables") have been the main contributors to the recent significant acceleration experienced by the share of RES (from 7% in 2007 to 9.6% in 2015).

8. Conclusions

An analysis of the EROI dynamics unequivocally suggests declining average EROIs for all fossil fuels [40], with the EROI of oil having likely halved in the short course of the first 15 years of the 21$^{st}$ century [13, 26, 46]. Consequently, the overall increase in oil production needed to keep pace with natural wealth and global population trajectories, that in this study is assessed to be >32 million oil b/d in 2025 over the current output, could even be affected by underestimation.

On the contrary, the EROI of solar PV energy is experiencing a strong increase, due to more than double in 2020 in comparison to 2010 [39]. Hence, previous assessments of energy and capital investments needed to achieve replacement of the oil fraction in the energy mix could be affected by overestimation [40, 46].

While the intrinsic inertia of the complex energy-economy system works against any fast paradigm shift in energy generation and use, projecting replacement of oil with RES in the energy mix into several decades to come [40], evidence exists that locally much faster transitions have occurred [47], including experiences with renewable energy sources in highly industrialized countries like Germany and Italy, where the renewable electricity output in 2015 approached, respectively, 33% [48] and 32.8% [49] of domestic power demand.

Driven by low cost, quickly improving EROI, and broad social acceptability arising from their concomitant ecofriendly nature and positive impact of growing penetration on the prices formed in the wholesale electricity market [50, 51], renewable energy sources will continue their penetration in the energy mix of developed and developing countries at fast pace, until low cost storage technology such as enzymatic hydrogen and $CO_2$ storage in renewable carbohydrates of high energy density [52] will become widely available, making the solar economy a common reality. Solar energy is mankind's common energy future, and its massive utilization through large-scale adoption of PV solar and wind electricity should be a priority for any Government across the world based on economic arguments such as those outlined in this study, prior to having to face the undesirable consequences of conflicting oil availability and population and wealth growth.

## Acknowledgments

This study is dedicated to Professor Y.-H. Percival Zhang, Virgina Tech, Tianjin Institute of Industrial Biotechnology, on the occasion of his lecture at Palermo's SuNEC 2016, for all he has done to advance the renewable carbohydrate electricity storage biotechnology.

**List of Abbreviations**

BP = British Petroleum

EIA = Energy Information Administration

EU = European Union

EROI = Energy Return On energy Invested

GDP = Gross Domestic Product

JODI = Joint Organizations Data Initiative

MTOE = Million Tons of Oil Equivalent

PV = Photovoltaic

RES = Renewable Energy Sources

ST = Solar Thermal

TEC = Total Energy Consumption

UN = United Nations

US = United States